# Conductance Model for Single-Crystalline/Compact Metal Oxide Gas Sensing Layers in the Non-Degenerate Limit: Example of Epitaxial SnO$_2$(101)

C. Simion[1], Federico Schipani[2], Alexandra Papadogianni[3], Adelina Stanoiu[1], Melanie Budde[3], Alexandru Oprea[2], Udo Weimar[2], Oliver Bierwagen[3], Nicolae Barsan*[2]

[1]National Institute of Materials Physics, Atomistilor 405A, 077125, Magurele, Romania

[2]Institute of Physical and Theoretical Chemistry, University of Tübingen, Auf der Morgenstelle 15, 72076 Tübingen, Germany

[3]Paul-Drude-Institute für Festkörperelektronik, Leibniz-Institut im Forschungsverbund Berlin e.V., Hausvogteiplatz 5–7, 10117 Berlin, Germany

## Abstract

Semiconducting metal oxide (SMOX)-based gas sensors are indispensable for safety and health applications, e.g. explosive, toxic gas alarms, controls for intake into car cabins and monitor for industrial processes.

In the past, the sensor community has been studying polycrystalline materials as sensors where the porous and random microstructure of the SMOX does not allow a separation of the phenomena involved in the sensing process. This lead to conduction models that can model and predict the behavior of the overall response, but they were not capable of giving fundamental information regarding the basic mechanisms taking place. The study of epitaxial layers is the definite prove to clarify the different aspects and contributions of the sensing mechanisms that are not possible to do by studying a polycrystalline sample.

A detailed analytical model for n and p-type single-crystalline/compact metal oxide gas sensors was developed that directly relates the conductance of the sample with changes in the surface electrostatic potential. Combined DC resistance and work function measurements were used in a compact SnO2 (101) layer in operando conditions that allowed us to check the validity of our model in the region where Boltzmann approximation holds to determine surface and bulk properties of the material.

*Corresponding author: nb@ipc.uni-tuebingen.de (N. Barsan).

More than 60 years ago Heiland et al. [1, 2] showed that the conductivity of semiconducting metal oxides (SMOX) (using single crystalline ZnO) depends on the composition of the surrounding atmosphere. More than 50 years ago Seiyama et al. [3] proposed the use of gas chromatographs equipped with SMOX – ZnO thin film – based detectors. The first gas sensors based on $SnO_2$, for the detection of natural gas leakages, were realized by Taguchi [4], who later founded Figaro Eng., the first company to commercialize SMOX based gas sensors in Japan. Since then, SMOX based sensors have been used in various applications e.g. explosive and toxic gas alarms, controls for air intake into car cabin, monitors for industrial processes etc. [5, 6]. Their technological realization was significantly changed and the combination between thick porous sensing layers and MEMS substrates, which offers significant miniaturization, cost reduction and lowering of the power consumption is now the state of the art [7, 8]. Recent developments are integrating different sensing materials on the same chip together with driving and evaluation electronics [9] heralding the occurrence of IoT ready SMOX-based sensing devices.

In principle, detection with SMOX based devices is simple; one generally assumes that: in air, at temperatures between 150 and 400 °C, atmospheric oxygen adsorbs at the surface of the metal oxides by trapping electrons from the bulk. This results in an overall increase of the sensor's resistance, for *n*-type materials, or a decrease, for *p*-type materials. The presence of gases in the atmosphere that react with the pre-adsorbed oxygen or directly with the metal oxide, also result in resistance changes (sensor signals). Accordingly, two aspects must be examined: the surface reaction taking place between the material and the gases together with the associated electrical charge transfer processes (sometimes called the receptor function) and their translation into the corresponding changes in the electrical resistance of the sensing layer (the transduction function)[10]; the latter plays a very important role e.g. possibly making the difference between the performances of *p*- and *n*-type SMOX based gas sensors [11]. For understanding the transduction function performing simultaneous work function and DC resistance measurements is extremely useful because it allows to directly link the change in the surface band bending with the sensor signals [12]. In combination with appropriate modeling this investigation method was proven to be very effective [13-17]. Most research effort, including modelling, was focused on porous sensing layers based on *n*-type SMOX because this combination is offering the best performance; the reason is that the transduction of the changes of the surface potential, band bending, are exponentially translated into changes of the electrical resistance of the sensing layer [18]. In the last period, there is a renewed interest for studying model systems, especially SMOX epitaxial layers [19, 20, 29]. Such studies answer the need to clarify different aspects of the fundamental understanding of sensing that are not possible to clarify by studying polycrystalline samples. For example, by combining operando IR studies and DFT calculations, it was found that the different ways in which water vapor influence the CO sensing of differently prepared $SnO_2$ sensing materials can be attributed to reaction taking place on different crystalline surfaces [21]. The definitive prove can be provided only by studying the corresponding epitaxial layers. Accordingly, here we developed a detailed model for the relationship between the conductance of compact *n* and *p*-type SMOX single crystalline films and the changes in the surface electrostatic potential determined by the charging of the surface with positive or negative charge. The model is restricted to cases in which the only charge carriers are electrons or holes that are obeying the Boltzmann statistics and there is no exchange of matter between the atmosphere and the bulk of the layer. In order to check its applicability, we realized and characterized epitaxial $SnO_2$ layers. The results of the combined work function/DC resistance measurements were used as inputs for calculation based on the model and allowed for the determination of surface and bulk properties of the material.

**Theoretical model**

The total conductance of a compact layer that is having its surface exposed to the ambient atmosphere is the sum between the conductance of the part of the layer influenced by the surface phenomena (called surface layer) and the conductance of the layer that is left unchanged (called bulk); we will call those: surface and volume contributions.

$$G_{total} = G_s + G_b \tag{1}$$

More in detail, equation 1 becomes:

$$G_{total} = \tilde{\sigma}_s \frac{z_0 W}{L} + \sigma_b \frac{(D-z_0)W}{L} \qquad (2)$$

where $L$ is the length of the layer, $W$ its width, $D$ its thickness and $z_0$ the thickness of the surface layer. $\tilde{\sigma}_s$ is the average conductivity of the surface layer and $\sigma_b$ is the conductivity of the bulk. For a more thorough analysis we will look to the possible specific cases. For the case of n-type SMOX the calculations are presented in detail. Because of the similarities, for the case of p-type SMOX only the final results are presented.

### *n*-type SMOX – Surface effects that do not affect the full layer

For an n-type SMOX, using the definition for conductivity, $\tilde{\sigma}_s = e\mu\tilde{n}_s$ and $\sigma_b = e\mu n_b$, equation 2 can be written:

$$G_{total} = \frac{e\mu\tilde{n}_s z_0 W}{L} + \frac{e\mu n_b (D-z_0)W}{L} = \frac{e\mu W}{L}[\tilde{n}_s z_0 + n_b(D-z_0)] =$$

$$= \frac{e\mu W}{L}[z_0(\tilde{n}_s - n_b) + n_b D] \qquad (3)$$

This case is described by the situation depicted in Figure 1.

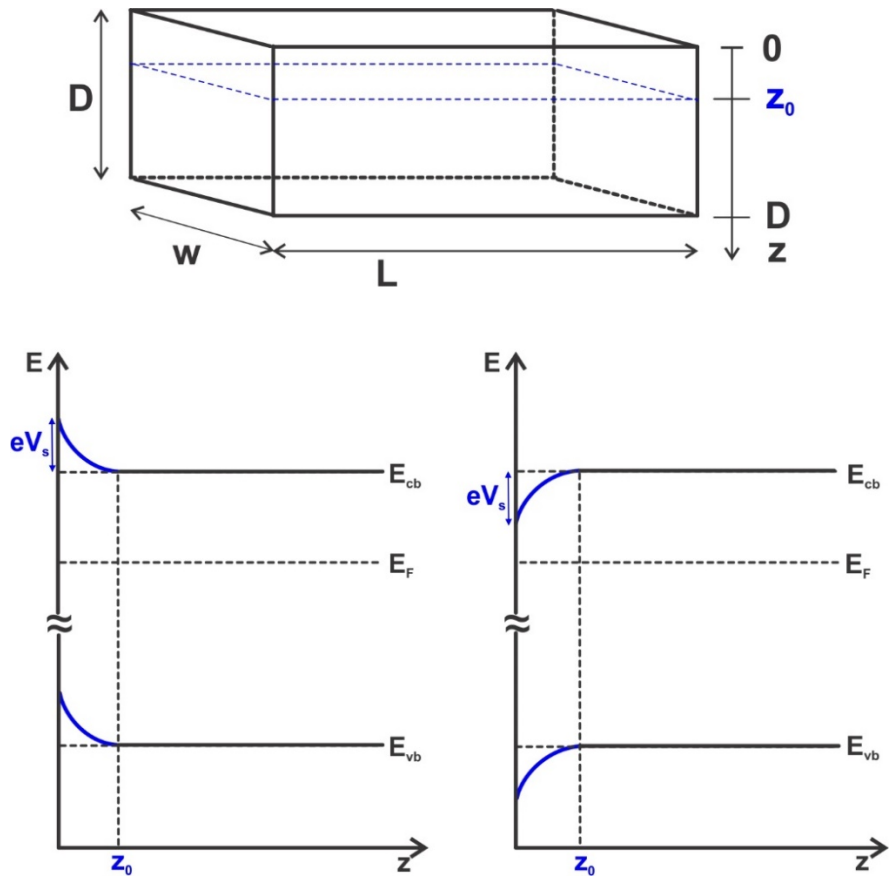

**Figure 1:** (up) Representation of a compact layer and ~~the~~ its geometrical characteristics: w-width; L-length; D-thickness; $z_0$-thickness of the space charge layer; (down left) Representation of thickness dependence of the energy bands in the semiconductor for the case of negative surface charge/formation of a depletion layer; (down right) Representation of thickness dependence of the energy bands in the semiconductor for the case of positive surface charge/formation of an accumulation layer.

In Figure 1 $V_S$ is defined as the difference between the electrostatic potential in the bulk and at the surface and $eV_S$ represents the corresponding energy difference. In the case of negative surface charge, the electrons will be repelled from the surface, which means that they will need more energy to get there. This is described in the energy band representation by an upward band bending and, correspondingly, the region close to the surface will have a lower density of electrons – a surface depletion layer is formed. Its extent, $z_0$, depends on the surface charge and the properties of the respective SMOX. The opposite happens in the case of positive surface charge, case in which a surface accumulation layer is formed. Detailed explanations are provided in [22].

Assuming the validity of the Boltzmann statistics in the whole layer and by observing that the only relevant change of properties is in direction $z$, the average electron concentration can be written as:

$$\widetilde{n_s} = \frac{1}{z_0} \int_0^{z_0} n_b \exp\left(-\frac{eV(z)}{k_B T}\right) dz \qquad (4)$$

It follows that:

$$\widetilde{n_s} - n_b = \frac{1}{z_0} \int_0^{z_0} n_b \exp\left(-\frac{eV(z)}{k_B T}\right) dz - \frac{1}{z_0} \int_0^{z_0} n_b \, dz \qquad (5)$$

$$\widetilde{n_s} - n_b = \frac{n_b}{z_0} \int_0^{z_0} \left[\exp\left(-\frac{eV(z)}{k_B T}\right) - 1\right] dz \qquad (6)$$

The dependence of thickness, z, is implicit through the dependence of band bending V. Changing the variables in equation 6 leads to:

$$\widetilde{n_s} - n_b = \frac{n_b}{z_0} \int_{V_s}^0 \left[\exp\left(-\frac{eV}{k_B T}\right) - 1\right] \left(\frac{dz}{dV}\right) dV = \frac{n_b}{z_0} \int_{V_s}^0 \frac{\exp\left(-\frac{eV}{k_B T}\right) - 1}{\left(\frac{dV}{dz}\right)} dV \qquad (7)$$

The general formula for $\left(\frac{dV}{dz}\right)$ is already available in the book of S. R. Morrison [22]. If one only takes into account the contribution of the electrons it takes the form:

$$\left(\frac{dV}{dz}\right) = \pm \left[\frac{2n_b k_B T}{\varepsilon \varepsilon_0}\right]^{1/2} \left[\exp\left(-\frac{eV}{k_B T}\right) + \frac{eV}{k_B T} - 1\right]^{1/2} \tag{8}$$

Combining equation 7 and equation 8 one obtains:

$$\widetilde{n_s} - n_b = \pm \frac{n_b}{z_0} \int_{V_S}^0 \frac{\exp\left(-\frac{eV}{k_B T}\right) - 1}{\left[\frac{2n_b k_B T}{\varepsilon \varepsilon_0}\right]^{\frac{1}{2}} \left[\exp\left(-\frac{eV}{k_B T}\right) + \frac{eV}{k_B T} - 1\right]^{\frac{1}{2}}} dV =$$

$$= \pm \frac{n_b}{z_0} \left[\frac{\varepsilon \varepsilon_0}{2n_b k_B T}\right]^{1/2} \int_{V_S}^0 \frac{\exp\left(-\frac{eV}{k_B T}\right) - 1}{\left[\exp\left(-\frac{eV}{k_B T}\right) + \frac{eV}{k_B T} - 1\right]^{1/2}} dV \tag{9}$$

The integral in equation 9 can be solved. In a first step we change the variable:

$$\frac{eV}{k_B T} = a, \; da = \frac{e}{k_B T} dV \tag{10}$$

which leads to:

$$\widetilde{n_s} - n_b = \pm \left[\frac{\varepsilon \varepsilon_0 n_b}{2 k_B T z_0^2}\right]^{1/2} \frac{k_B T}{e} \int_{\frac{eV_S}{k_B T}}^0 \frac{\exp(-a) - 1}{[\exp(-a) + a - 1]^{1/2}} da \tag{11}$$

and to a subsequent change of variable:

$$\exp(-a) + a - 1 = m \tag{12}$$

Which, with the observation that:

$$dm = [-\exp(-a) + 1] da = -[\exp(-a) - 1] da \tag{13}$$

brings us to:

$$\int_{\frac{eV_S}{k_B T}}^0 \frac{\exp(-a) - 1}{[\exp(-a) + a - 1]^{1/2}} da = -\int_m^0 \frac{1}{m^{\frac{1}{2}}} dm = \int_0^m \frac{1}{m^{\frac{1}{2}}} dm =$$

$$= 2\sqrt{m}\Big|_0^m \xrightarrow{yields} 2 \left[\exp\left(-\frac{eV_s}{k_B T}\right) + \frac{eV_s}{k_B T} - 1\right]^{1/2} \tag{14}$$

Using equation 14, equation 11 can be written as:

$$\widetilde{n_s} - n_b = \pm \left[\frac{\varepsilon \varepsilon_0 n_b}{2 k_B T z_0^2}\right]^{\frac{1}{2}} 2 \frac{k_B T}{e} \left[\exp\left(-\frac{eV_s}{k_B T}\right) + \frac{eV_s}{k_B T} - 1\right]^{\frac{1}{2}} =$$

$$= \pm \frac{n_b}{z_0} \left[\frac{4\varepsilon \varepsilon_0 k_B^2 T^2}{2 n_b k_B T e^2}\right]^{1/2} \left[\exp\left(-\frac{eV_s}{k_B T}\right) + \frac{eV_s}{k_B T} - 1\right]^{1/2} =$$

$$= \pm \frac{n_b}{z_0} \sqrt{2} \left[\frac{\varepsilon \varepsilon_0 k_B T}{n_b e^2}\right]^{1/2} \left[\exp\left(-\frac{eV_s}{k_B T}\right) + \frac{eV_s}{k_B T} - 1\right]^{1/2} \tag{15}$$

Knowing that the Debye length, $L_D$ is defined as $L_D = \left[\frac{\varepsilon \varepsilon_0 k_B T}{n_b e^2}\right]^{1/2}$ one gets:

$$\widetilde{n_s} - n_b = \pm \frac{n_b}{z_0} \sqrt{2} L_D \left[ \exp\left(-\frac{eV_s}{k_B T}\right) + \frac{eV_s}{k_B T} - 1 \right]^{1/2} \qquad (16)$$

Replacing equation 16 into equation 3 one obtains for the conductance of a non-fully depleted n-type SMOX layer:

$$G_{total} = \frac{e\mu W}{L} \left\{ \pm n_b \sqrt{2} L_D \left[ \exp\left(-\frac{eV_s}{k_B T}\right) + \frac{eV_s}{k_B T} - 1 \right]^{1/2} + n_b D \right\} =$$

$$= \frac{e\mu W n_b D}{L} \left\{ \pm \sqrt{2} \frac{L_D}{D} \left[ \exp\left(-\frac{eV_s}{k_B T}\right) + \frac{eV_s}{k_B T} - 1 \right]^{1/2} + 1 \right\} \qquad (17)$$

The formula can be further simplified by observing that one can introduce the bulk conductance of the whole layer as $G_b = \frac{e\mu W n_b D}{L}$ into equation 17. It results that the dependence of the overall conductance on the surface potential (band-bending) is:

$$G_{total} = G_b \left[ 1 \pm \sqrt{2} \frac{L_D}{D} \left[ \exp\left(-\frac{eV_s}{k_B T}\right) + \frac{eV_s}{k_B T} - 1 \right]^{1/2} \right] \qquad (18)$$

There are two possibilities, $V_s > 0$, which indicates the formation of a depletion layer, and $V_s < 0$, which means the formation of an accumulation layer. In the first case the conductance will decrease so one would have to use – in equation 18; in the second case the conductance will increase so one would have to use + in equation 18.

**For the case of the depletion layer**, $V_s > 0$ equation 18 becomes:

$$G_{total} = G_b \left[ 1 - \sqrt{2} \frac{L_D}{D} \left[ \exp\left(-\frac{eV_s}{k_B T}\right) + \frac{eV_s}{k_B T} - 1 \right]^{1/2} \right] \qquad (19)$$

In case of very small band bendings when compared to the thermal energy, $\frac{e|V_s|}{k_B T} \ll 1$ one can use a Taylor series approximation for $\exp\left(-\frac{eV_s}{k_B T}\right)$ and keep just the first three terms:

$$\left[ \exp\left(-\frac{eV_s}{k_B T}\right) + \frac{eV_s}{k_B T} - 1 \right]^{\frac{1}{2}} \cong \left[ 1 - \frac{eV_s}{k_B T} + \frac{1}{2}\left(\frac{eV_s}{k_B T}\right)^2 + \cdots + \frac{eV_s}{k_B T} - 1 \right]^{1/2} \cong \frac{1}{\sqrt{2}} \left(\frac{eV_s}{k_B T}\right) \qquad (20)$$

Equation 19 becomes:

$$G_{total} = G_b \left[ 1 - \frac{L_D}{D} \left(\frac{eV_s}{k_B T}\right) \right] \qquad (21)$$

which indicates a linear dependency on $V_S$ around 0.

In case of very large band bendings when compared to the thermal energy, $\frac{e|V_s|}{k_B T} \gg 1$,

$$\left[ \exp\left(-\frac{eV_s}{k_B T}\right) + \frac{eV_s}{k_B T} - 1 \right]^{\frac{1}{2}} \cong \left(\frac{eV_s}{k_B T}\right)^{\frac{1}{2}} \qquad (22)$$

equation 19 becomes

$$G_{total} = G_b \left[1 - \sqrt{2}\frac{L_D}{D}\left(\frac{eV_s}{k_BT}\right)^{1/2}\right] \quad (23)$$

This result suggests a linear dependence of the conductance on the square root of the band bending.

On the basis of equation 23 we can also identify the limit of validity for the formula described by equation 19, namely

$$\sqrt{2}\frac{L_D}{D}\left(\frac{eV_s}{k_BT}\right)^{1/2} \leq 1 \rightarrow \frac{eV_s}{k_BT} \leq \left(\frac{D}{\sqrt{2}L_D}\right)^2 \rightarrow eV_s \leq \frac{k_BT}{2}\left(\frac{D}{L_D}\right)^2; \quad (24)$$

At larger values of the band bending, the depletion layer extends in the full layer and one needs a different approach.

**For the case of the accumulation layer**, $V_s < 0$ equation 18 becomes:

$$G_{total} = G_b \left[1 + \sqrt{2}\frac{L_D}{D}\left[exp\left(-\frac{eV_s}{k_BT}\right) + \frac{eV_s}{k_BT} - 1\right]^{1/2}\right] \quad (25)$$

At low values of the band bending one obtains:

$$G_{total} = G_b \left[1 + \frac{L_D}{D}\left(\frac{e|V_s|}{k_BT}\right)\right] \quad (26)$$

For large values of the band bending,

$$\left[exp\left(-\frac{eV_s}{k_BT}\right) + \frac{eV_s}{k_BT} - 1\right] \cong exp\left(-\frac{eV_s}{k_BT}\right) = exp\left(\frac{e|V_s|}{k_BT}\right); \quad (27)$$

then equation 25 became:

$$G_{total} = G_b \left[1 + \sqrt{2}\frac{L_D}{D}\left(exp\frac{e|V_s|}{k_BT}\right)^{1/2}\right] = G_b \left[1 + \sqrt{2}\frac{L_D}{D}\left(exp\frac{e|V_s|}{2k_BT}\right)\right] \quad (28)$$

In extreme cases $G_{total} \sim exp\frac{e|V_s|}{2k_BT}$ as in the case of a conduction mechanism dominated by the surface accumulation layer for porous layers [23], which indicates that the conduction taking place through the accumulation layer dominates.

Figure 2 presents the relationship between the normalized conduction of the compact layer $\frac{G}{G_b}$ as a function of the band bending expressed in *kT* units for the case of an *n*-type SMOX. Three cases are presented, namely Debye length representing 1%, 20% and 50% from the total layer thickness.

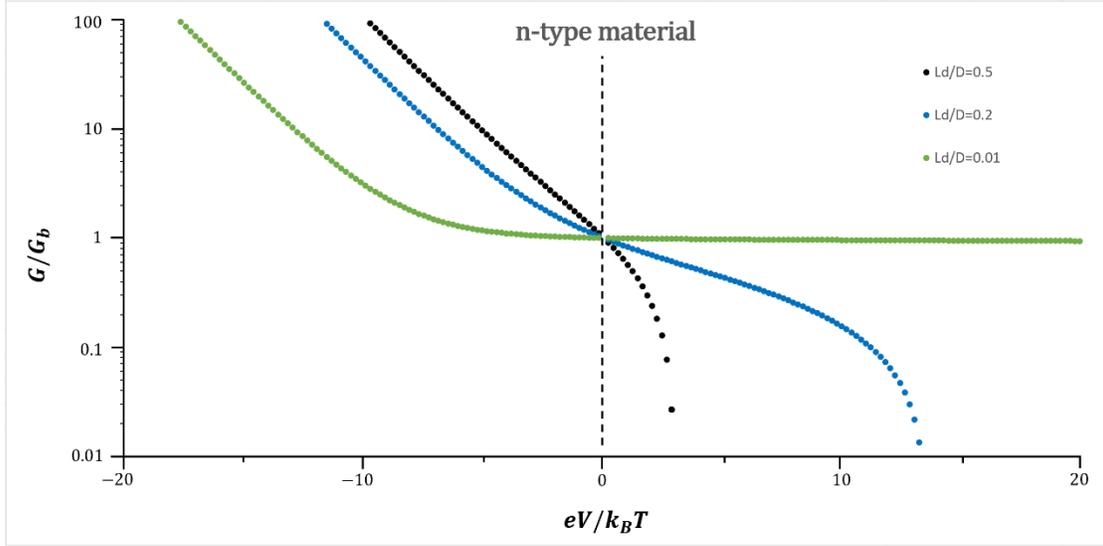

**Figure 2: Normalized conduction of a n-type SMOX sensing layer as a function of surface band bending for three different values of $\frac{L_D}{D}$; negative values of the band bending correspond to the formation of a surface accumulation layer while positive values of the band bending correspond to the formation of a surface depletion layer.**

In Figure 2, for the case of positive band bending and $\frac{L_D}{D}$=0.5, the validity of the model ends, i. e.:

$$\sqrt{2}\frac{L_D}{D}\left(\frac{eV_S}{k_BT}\right)^{1/2} = 1 \tag{29}$$

already at $\frac{eV_S}{k_BT} \cong 2$. At higher values of the band bending the depletion layer fully extends in the whole layer. This case is presented in the following section.

## n-type SMOX – Surface effects that do affect the full layer

This situation is presented in Figure 3.

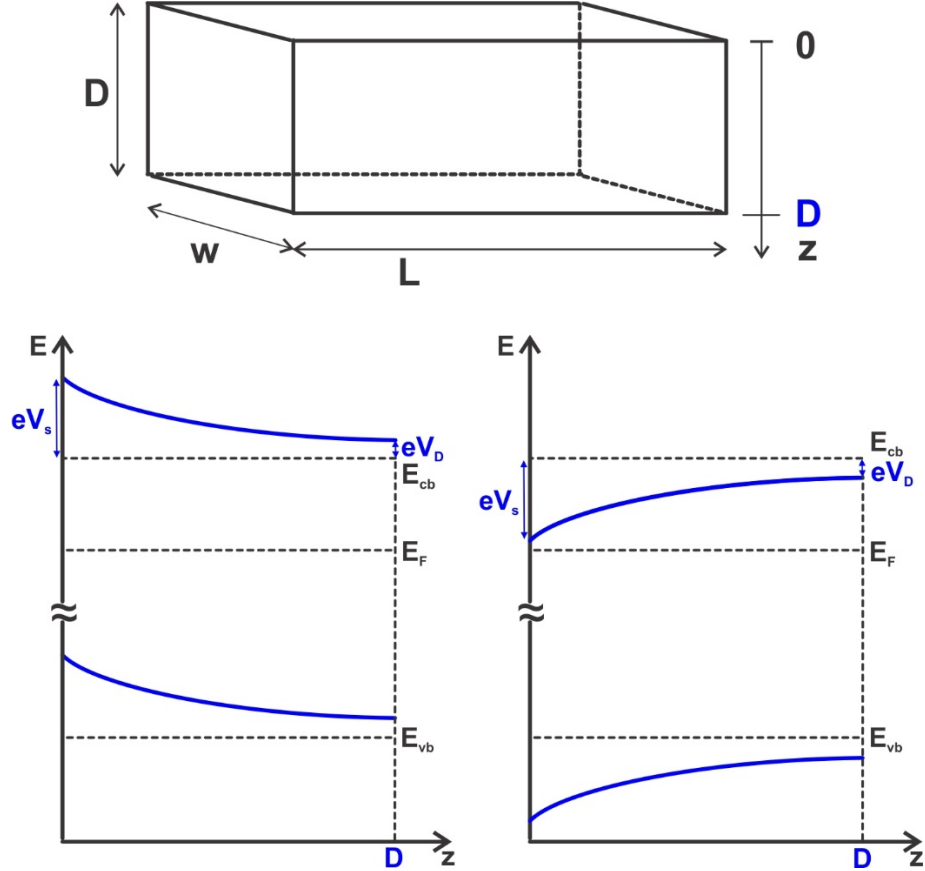

**Figure 3: (up) Representation of a compact layer and its geometrical characteristics: w-width; L-length; D-thickness; z0-thickness of the space charge layer; (down left) Representation of thickness dependence of the energy bands in the semiconductor for the case of negative surface charge/formation of a depletion layer; (down right) Representation of thickness dependence of the energy bands in the semiconductor for the case of positive surface charge/formation of an accumulation layer.**

In this case one can write for the conductance:

$$G_{total} = \frac{e\mu W}{L} \tilde{n} D \tag{30}$$

The average electron concentration can be expressed by:

$$\tilde{n} = \frac{1}{D}\int_0^D n_b \exp\left(-\frac{eV(z)}{k_B T}\right) dz \tag{31}$$

In order to follow approach for the integration described be equations 9 to equation 17 one needs to form again the term $\frac{n_b}{z_0}\int_0^{z_0}\left[\exp\left(-\frac{eV(z)}{k_BT}\right)-1\right]dz$. To do so one can add and subtract $n_b$:

$$\tilde{n}-n_b+n_b = \frac{1}{D}\int_0^D n_b \exp\left(-\frac{eV(z)}{k_BT}\right)dz - \frac{1}{D}\int_0^D n_b\,dz + n_b = n_b + \frac{n_b}{D}\int_0^D\left[\exp\left(-\frac{eV(z)}{k_BT}\right)-1\right]dz \quad (32)$$

By changing the variable, using equation 8 and applying the boundary conditions:

$z = 0 \rightarrow V(z) = V_S$

$z = D \rightarrow V(z) = V_D$

One obtains:

$$G_{total} = G_b\left\{1 \pm \sqrt{2}\frac{L_D}{D}\left[\left[\exp\left(-\frac{eV_S}{k_BT}\right)+\frac{eV_S}{k_BT}-1\right]^{1/2} - \left[\exp\left(-\frac{eV_D}{k_BT}\right)+\frac{eV_D}{k_BT}-1\right]^{1/2}\right]\right\} \quad (33)$$

**For the case of the depletion layer**, $V_S > 0$, equation 33 becomes:

$$G_{total} = G_b\left\{1 - \sqrt{2}\frac{L_D}{D}\left[\left[\exp\left(-\frac{eV_S}{k_BT}\right)+\frac{eV_S}{k_BT}-1\right]^{1/2} - \left[\exp\left(-\frac{eV_D}{k_BT}\right)+\frac{eV_D}{k_BT}-1\right]^{1/2}\right]\right\} \quad (34)$$

For large values of the surface band bending throughout the full layer, both $\frac{eV_S}{k_BT}$ and $\frac{eV_D}{k_BT} \gg 1$ so one can approximate:

$$\left[\exp\left(-\frac{eV_S}{k_BT}\right)+\frac{eV_S}{k_BT}-1\right]^{1/2} \cong \left(\frac{eV_S}{k_BT}\right)^{1/2} \quad (35)$$

$$\left[\exp\left(-\frac{eV_D}{k_BT}\right)+\frac{eV_D}{k_BT}-1\right]^{1/2} \cong \left(\frac{eV_D}{k_BT}\right)^{1/2} \quad (36)$$

Thus, equation 34 becomes:

$$G_{total} \cong G_b\left\{1 - \sqrt{2}\frac{L_D}{D}\left[\left(\frac{eV_S}{k_BT}\right)^{\frac{1}{2}} - \left(\frac{eV_D}{k_BT}\right)^{\frac{1}{2}}\right]\right\} \quad (37)$$

In the same conditions, equation 8 can be simplified, namely:

$$\left(\frac{dV}{dz}\right) = \pm\left[\frac{2n_bk_BT}{\varepsilon\varepsilon_0}\right]^{1/2}\left[\exp\left(-\frac{eV}{k_BT}\right)+\frac{eV}{k_BT}-1\right]^{1/2} \cong \pm\left(\frac{2n_bk_BT}{\varepsilon\varepsilon_0}\right)^{1/2}\left(\frac{eV}{k_BT}\right)^{1/2} \quad (38)$$

It can also be integrated after separating the variables:

$$\frac{dV}{\left(\frac{eV}{k_BT}\right)^{1/2}} = \pm\left(\frac{2n_bk_BT}{\varepsilon\varepsilon_0}\right)^{1/2}dz = \pm\sqrt{2}\frac{k_BT}{L_De}dz \quad (39)$$

The result of the integration is:

$$\left[\left(\frac{eV_S}{k_BT}\right)^{1/2} - \left(\frac{eV_D}{k_BT}\right)^{1/2}\right] = \frac{D}{\sqrt{2}L_D} \quad (40)$$

Equation 37 becomes:

$$G_{total} \cong G_b\left(1 - \sqrt{2}\frac{L_D}{D}\frac{D}{\sqrt{2}L_D}\right) \cong G_b(1-1) \cong 0, \quad (41)$$

That can be interpreted as an indication that the layer will become insulating - electrical resistance $R \to \infty$.

**For the case of the accumulation layer**, $V_s < 0$, equation 33 becomes:

$$G_{total} = G_b \left\{ 1 + \sqrt{2} \frac{L_D}{D} \left[ \left[ \exp\left(-\frac{eV_s}{k_BT}\right) + \frac{eV_s}{k_BT} - 1 \right]^{1/2} - \left[ \exp\left(-\frac{eV_D}{k_BT}\right) + \frac{eV_D}{k_BT} - 1 \right]^{1/2} \right] \right\} \quad (42)$$

For large values of the surface band bending: $\frac{e|V_s|}{k_BT} \gg 1$ and $\frac{e|V_D|}{k_BT} \gg 1$ one can simplify equation 42 by observing that:

$$\left[ \exp\left(-\frac{eV_s}{k_BT}\right) + \frac{eV_s}{k_BT} - 1 \right]^{1/2} \cong \left( \exp \frac{e|V_s|}{k_BT} \right)^{1/2} \quad (43)$$

$$\left[ \exp\left(-\frac{eV_D}{k_BT}\right) + \frac{eV_D}{k_BT} - 1 \right]^{1/2} \cong \left( \exp \frac{e|V_D|}{k_BT} \right)^{1/2} \quad (44)$$

Thus, equation 42 can be written as:

$$G_{total} \cong G_b \left[ 1 + \sqrt{2} \frac{L_D}{D} \left( \exp \frac{e|V_s|}{2k_BT} - \exp \frac{e|V_D|}{2k_BT} \right) \right] \quad (45)$$

Equation 8:

$$\left(\frac{dV}{dz}\right) = \pm \left[ \frac{2n_b k_B T}{\varepsilon \varepsilon_0} \right]^{1/2} \left[ \exp\left(-\frac{eV}{k_BT}\right) + \frac{eV}{k_BT} - 1 \right]^{1/2} \quad (8)$$

becomes:

$$\frac{dV}{dz} = \pm \left( \frac{2n_b k_B T}{\varepsilon \varepsilon_0} \right)^{1/2} \left( \exp \frac{e|V|}{k_BT} \right)^{1/2} \quad (46)$$

It can also be integrated after separating the variables:

$$\frac{dV}{\exp \frac{e|V|}{2k_BT}} = \pm \left( \frac{2n_b k_B T}{\varepsilon \varepsilon_0} \right)^{1/2} dz = \pm \sqrt{2} \frac{k_BT}{L_D e} dz \quad (47)$$

Integrating one obtains:

$$\left( \exp \frac{e|V_s|}{2k_BT} - \exp \frac{e|V_D|}{2k_BT} \right) = \sqrt{2} \frac{D}{L_D} \exp \frac{e|V_D| + e|V_s|}{2k_BT} \quad (48)$$

Equation 45 becomes:

$$G_{total} \cong G_b \left[ 1 + 2\exp \frac{e|V_D| + e|V_s|}{2k_BT} \right] \quad (49)$$

Equation 49 suggests that the conductance of an accumulation layer fully extended throughout the full sensing layer depends exponentially of some kind of average band bending. In the case of the flat band, $e|V_D| = e|V_s|$ so one obtains

$$G_{total} \cong G_b \left[ 1 + 2\exp \frac{e|V_s|}{k_BT} \right] \quad (50)$$

## *p-type SMOX – Surface effects that do not affect the full layer*

This case is described in Figure 4.

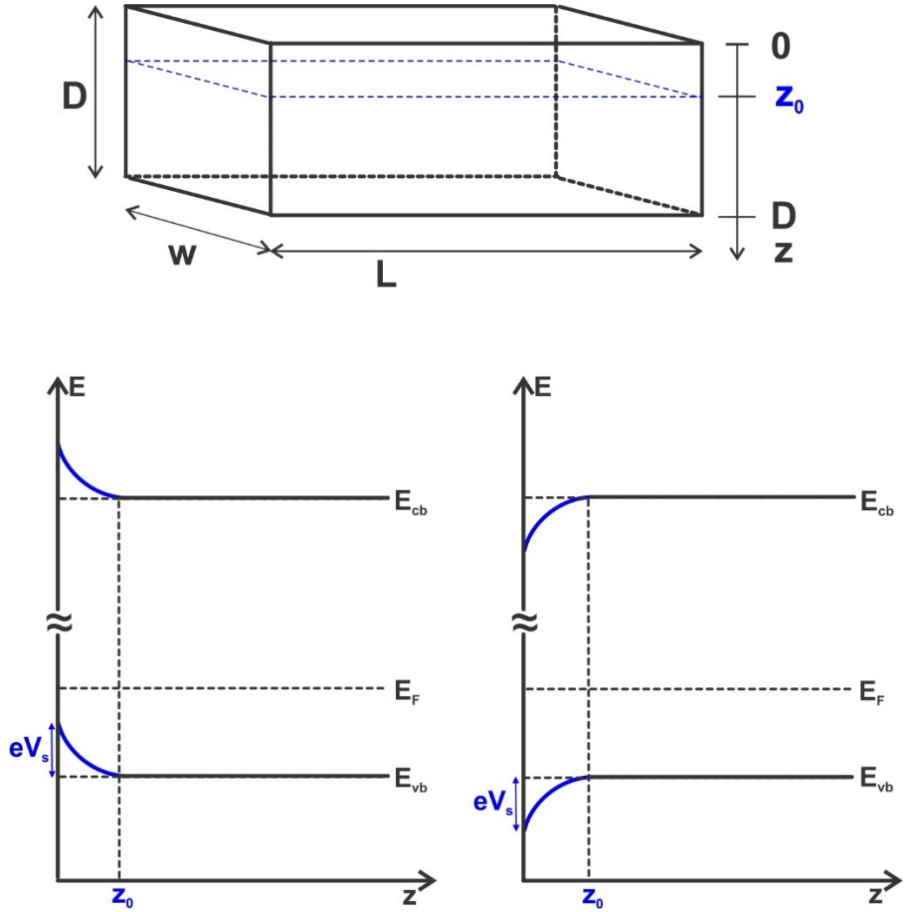

**Figure 4: (up) Representation of a compact layer and the its geometrical characteristics: w-width; L-length; D-thickness; z0-thickness of the space charge layer; (down left) Representation of thickness dependence of the energy bands in the semiconductor for the case of positive surface charge/formation of a accumulation layer; (down right) Representation of thickness dependence of the energy bands in the semiconductor for the case of negative surface charge/formation of an depletion layer.**

The general form of conductance for the compact *p*-type SMOX layer can be written as:

$$G_{total} = G_s + G_b \sim \tilde{p}_s z_0 + p_b(D - z_0) \qquad (51)$$

Where the term $\tilde{p}_s z_0$ represents the surface contribution and the term $p_b(D - z_0)$ represents the bulk contribution.

Following the same mathematical approach as in the case of n-type SMOX, one finally obtains:

$$G_{total} = G_b \left[1 \pm \sqrt{2}\frac{L_D}{D}\left[\exp\left(\frac{eV_s}{k_BT}\right) - \frac{eV_s}{k_BT} - 1\right]^{1/2}\right] \quad (52)$$

There are two possibilities, $V_s > 0$, which indicates the formation of an accumulation layer and $V_s < 0$, which means the formation of a depletion layer. In the first case the conductance will increase so one would have to use + in equation 52; in the second case the conductance will decrease therefore one would have to use – in equation 52.

Figure 5 presents the relationship between the normalized conduction of the compact layer $\frac{G}{G_b}$ as a function of the band bending expressed in $kT$ units for the case of an p-type SMOX. Three cases are presented, namely Debye length representing 1%, 20% and 50% from the total layer thickness.

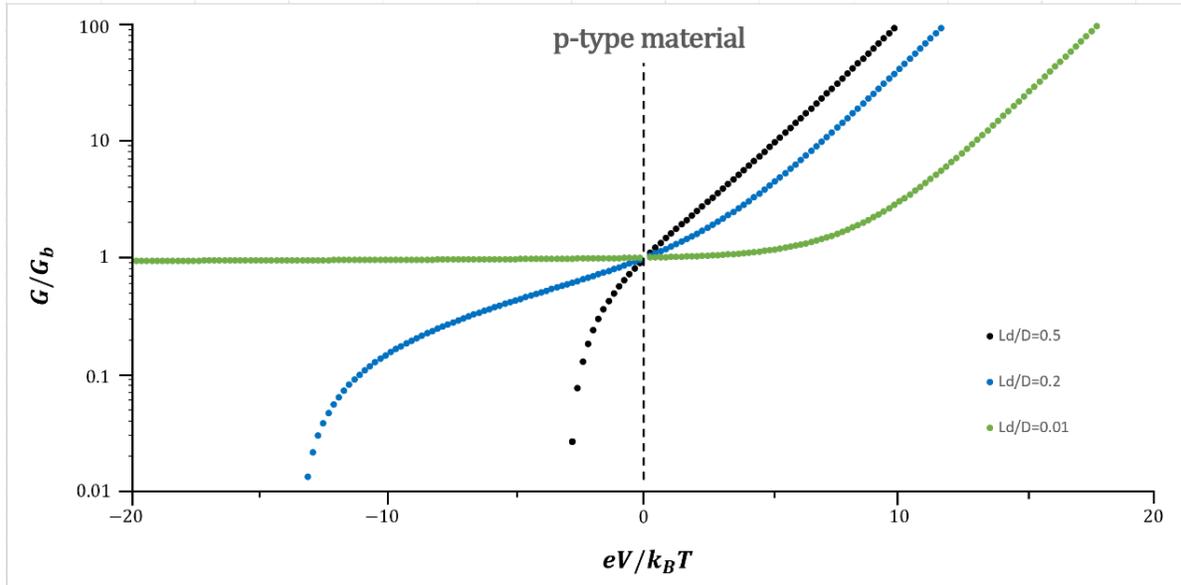

**Figure 5: Normalized conduction of a p-type SMOX sensing layer as a function of surface band bending for three different values of $\frac{L_D}{D}$; negative values of the band bending correspond to the formation of a surface depletion layer while positive values of the band bending correspond to the formation of a surface accumulation layer.**

As in the case of n-type SMOX one can obtain the limits of validity for the model, in this case for the negative band bending $-\sqrt{2}\frac{L_D}{D}\left(-\frac{eV_s}{k_BT}\right)^{1/2} = 1$ and also determine the approximations valid at very low and very large values of band bending.

**For the case of the accumulation layer**, $V_s > 0$ they are:

$$G_{total} = G_b \left[1 + \frac{L_D}{D}\left(\frac{eV_s}{k_BT}\right)\right] \tag{53}$$

in case of very small band bending when compared to the thermal energy, $\frac{eV_s}{k_BT} \ll 1$ and

$$G_{total} = G_b \left[1 + \sqrt{2}\frac{L_D}{D} exp\left(\frac{eV_s}{2k_BT}\right)\right] \tag{54}$$

in case of very large band bending when compared to the thermal energy, $\frac{eV_s}{k_BT} \gg 1$.

**For the case of the depletion layer**, $V_s < 0$ they are:

$$G_{total} = G_b \left[1 - \frac{L_D}{D}\left(\frac{e|V_s|}{k_BT}\right)\right] \tag{55}$$

in case of very small band bending when compared to the thermal energy, $\frac{e|V_s|}{k_BT} \ll 1$ and

$$G_{total} = G_b \left[1 - \sqrt{2}\frac{L_D}{D}\left(\frac{e|V_s|}{k_BT}\right)^{1/2}\right] \tag{56}$$

in case of very large band bending when compared to the thermal energy, $\frac{e|V_s|}{k_BT} \gg 1$.

## p-type SMOX – Surface influence extended in the full layer

This case is described in Figure 6

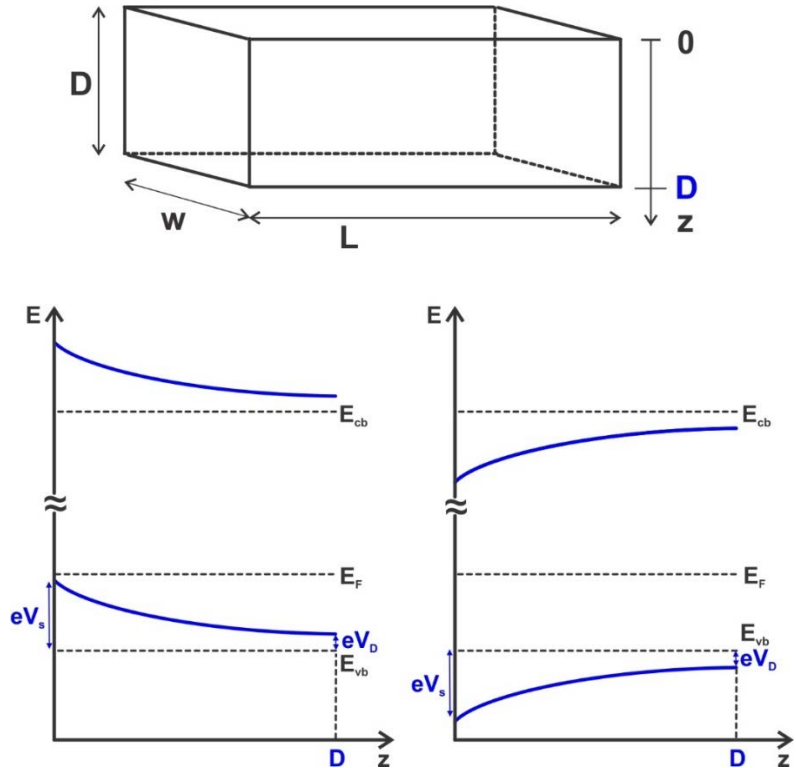

**Figure 6: (up) Representation of a compact layer and its geometrical characteristics: w-width; L-length; D-thickness; z0-thickness of the space charge layer; (down left) Representation of thickness**

dependence of the energy bands in the semiconductor for the case of negative surface charge/formation of an accumulation layer; (down right) Representation of thickness dependence of the energy bands in the semiconductor for the case of positive surface charge/formation of a depletion layer.

By using the same mathematical approach like in the case of n-type SMOX one obtains:

$$G_{total} = G_b \left\{ 1 \pm \sqrt{2} \frac{L_D}{D} \left[ \left[ exp\left(\frac{eV_s}{k_B T}\right) - \frac{eV_s}{k_B T} - 1 \right]^{1/2} - \left[ exp\left(\frac{eV_D}{k_B T}\right) - \frac{eV_D}{k_B T} - 1 \right]^{1/2} \right] \right\} \quad (57)$$

**For the case of the accumulation layer**, $V_s > 0$ equation 57 becomes:

$$G_{total} = G_b \left\{ 1 + \sqrt{2} \frac{L_D}{D} \left[ \left[ exp\left(\frac{eV_s}{k_B T}\right) - \frac{eV_s}{k_B T} - 1 \right]^{1/2} - \left[ exp\left(\frac{eV_D}{k_B T}\right) - \frac{eV_D}{k_B T} - 1 \right]^{1/2} \right] \right\} \quad (58)$$

For large values of the surface band bending throughout the full layer, both $\frac{eV_s}{k_B T}$ and $\frac{eV_D}{k_B T} \gg 1$ so one obtains (see the approach used in the case of n-type SMOX, equation 45 to equation 49):

$$G_{total} \cong G_b \left( 1 + 2exp\frac{eV_D + eV_s}{2k_B T} \right) \quad (59)$$

suggests that the conductance of an accumulation layer fully extended throughout the full sensing layer depends exponentially of some kind of average band bending. In the case of the flat band, $eV_D = eV_s$ so one obtains

$$G_{total} \cong G_b \left[ 1 + 2exp\frac{eV_s}{k_B T} \right] \quad (60)$$

**For the case of the depletion layer**, $V_s < 0$ becomes:

$$G_{total} = G_b \left\{ 1 - \sqrt{2} \frac{L_D}{D} \left[ \left[ exp\left(\frac{eV_s}{k_B T}\right) - \frac{eV_s}{k_B T} - 1 \right]^{1/2} - \left[ exp\left(\frac{eV_D}{k_B T}\right) - \frac{eV_D}{k_B T} - 1 \right]^{1/2} \right] \right\} \quad (61)$$

For large values of the surface band bending throughout the full layer (see the approach used in the case of n-type SMOX, equation 37 to equation 40) one obtains:

$$G_{total} \cong G_b \left\{ 1 - \sqrt{2} \frac{L_D}{D} \left[ \left(\frac{e|V_s|}{k_B T}\right)^{\frac{1}{2}} - \left(\frac{e|V_D|}{k_B T}\right)^{\frac{1}{2}} \right] \right\} \quad (62)$$

which, like in the case of *n*-type SMOX, will tend to 0.

## Experimental validation

### Sample preparation and characterization

A 280 nm-thick, single crystalline, unintentionally-doped, rutile SnO$_2$(1 0 1) film was grown on a quarter of a 2" diameter r-plane sapphire substrate by plasma-assisted molecular beam epitaxy similar to the work described in [32]. The rough backside of the substrate was coated by 1 µm Ti to facilitate radiative heating by a heating filament behind the substrate. The substrate temperature was measured by a thermocouple placed between heater and substrate. Activated oxygen was provided by passing a controlled flow of molecular oxygen (6N purity) through an RF-plasma source (SPECS, PCF-RF-AN) run at a power of 300 W. Highly pure Sn (7N purity) was evaporated from a single-filament effusion cell at 1170°C equipped with a shutter between cell and substrate. The resulting flux of Sn corresponded to a beam equivalent pressure of $3.7 \times 10^{-7}$ Torr as measured by a nude filament ion gauge in the substrate position. Prior to growth the substrate was exposed to an activated oxygen flux of 0.5 standard cubic centimeter per minute (sccm) at a substrate temperature of 850°C to improve the surface quality. After that the substrate temperature was decreased to 750°C and growth was initiated by opening the Sn shutter to grow a nucleation layer for 7 min. The substrate temperature was subsequently ramped up to 850°C within 100 s without interrupting the growth. After 26 min of growth the oxygen flux was increased to 1 sccm, which improved the surface smoothness gauged by reflection high energy electron beam diffraction (RHEED). Growth was terminated after a total growth time of 1 h by closing the Sn shutter, closing the oxygen flux, and cooling down the substrate in vacuum at 0.3°C/s. The total film thickness is approx. 280 nm measured by in-situ laser reflectometry during growth.

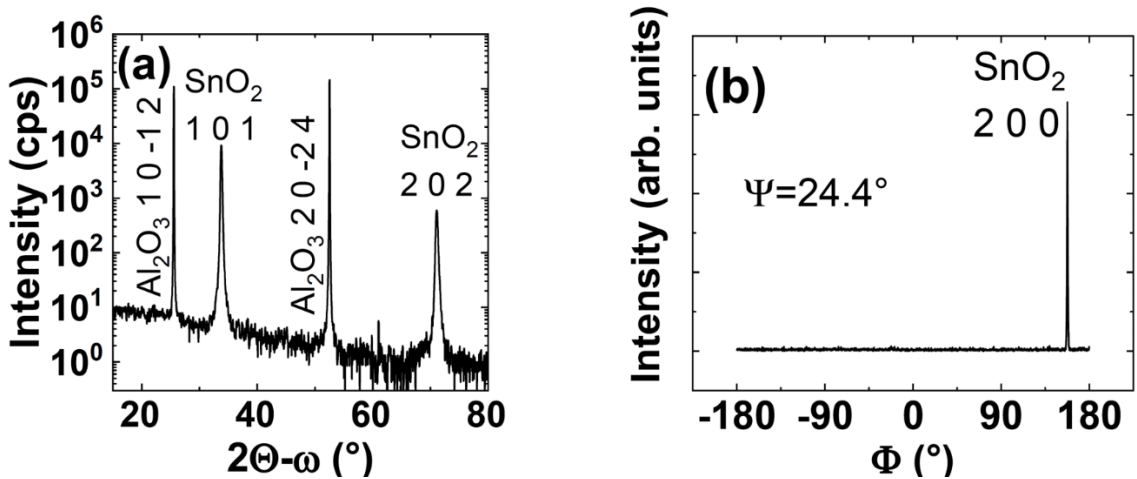

**Figure 7: XRD measurements of the grown film. (a) Symmetric out-of-plane 2Θ-ω scan. The diffraction peaks are labeled. (b) Φ-scan of the skew-symmetric 2 0 0 reflection with sample tilt angle Ψ.**

X-ray diffraction (XRD) confirmed the single crystallinity of the film with (1 0 1) out-of-plane orientation: The exclusive presence of diffraction orders of the $Al_2O_3$(1 0 -1 2) and $SnO_2$(1 0 1) planes in the symmetric out-of-plane 2Θ-ω scan of the film shown in Figure 7(a) indicates a phase-pure rutile $SnO_2$(1 0 1) film. The presence of a single peak in the Φ-scan of the skew-symmetric $SnO_2$(2 0 0) reflex measured by rotating the sample around its surface normal by the angle Φ, shown in Figure 7(b), indicates a single in-plane orientation without rotational domains.

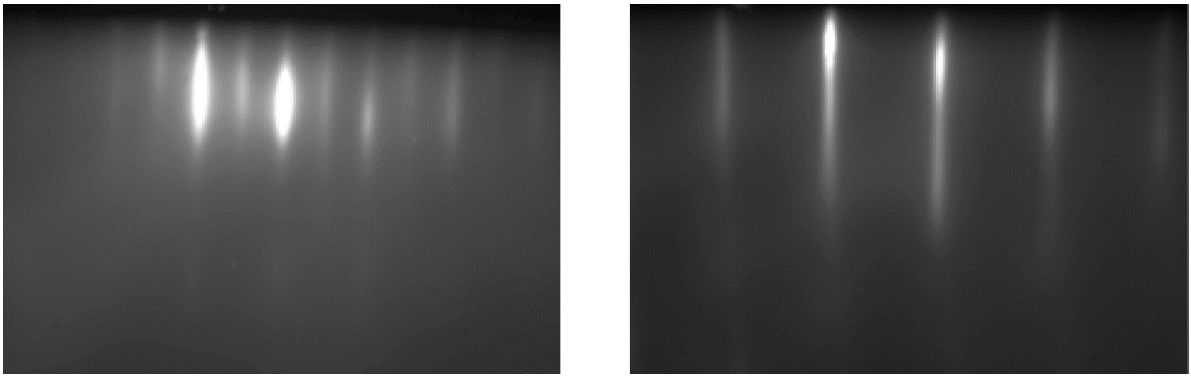

**Figure 8: RHEED pattern measured after growth in two perpendicular azimuths. The patterns consisting of vertical streaks and the absence of spots indicate a smooth, crystalline surface.**

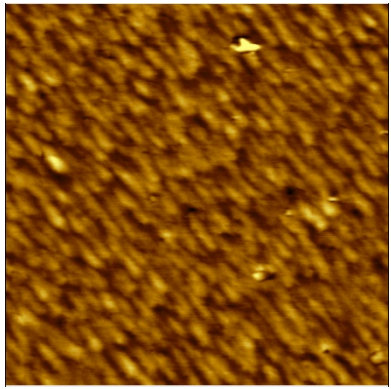

**Figure 9: 2 µm x 2 µm AFM image of the film surface. The height scale (black-to-white) is 3 nm.**

The RHEED patterns shown in Figure 8 and the peak-force-tapping mode atomic force microscopy (AFM) image shown in Figure 9 indicate a smooth, unfaceted $SnO_2$ surface.

After growth the quarter wafer was cleaved into rectangular chips. Van-der-Pauw measurements at room temperature and in ambient air (i.e. with humidity) of a 5 mm x 6 mm chip with In-contacts pressed on the film surface in the corners indicated a sheet resistance of 12 MΩ, corresponding to a conductance of 0.083 µS. For the gas-response measurements, a

4 mm x 7 mm chip was cleaved. Disk-shaped 20 nm Ti/100 nm Au contacts with 0.5 mm diameter were deposited by electron-beam evaporation through a shadow mask in a 2 mm x 2 mm square arrangement next to a short edge of the chip. This metallization provides an ohmic contact to $SnO_2$ [28].

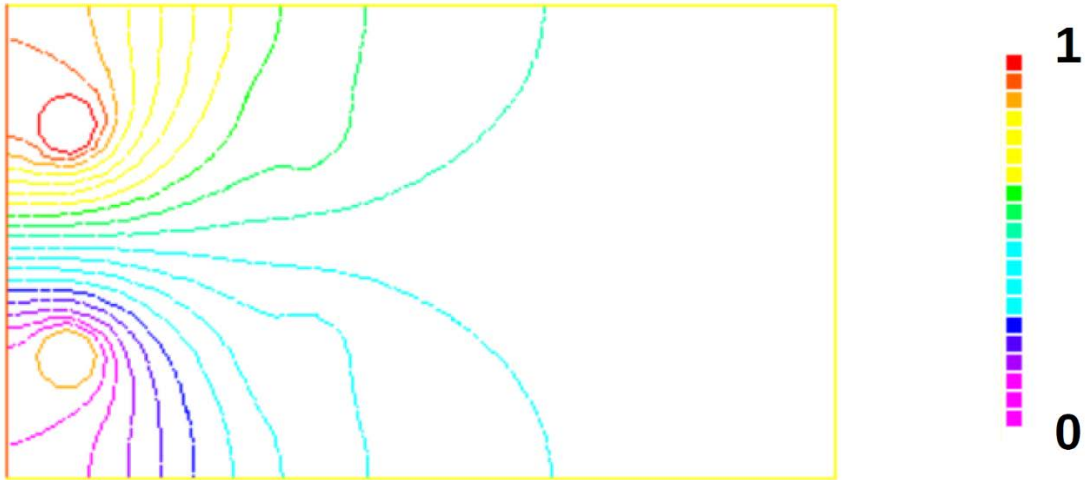

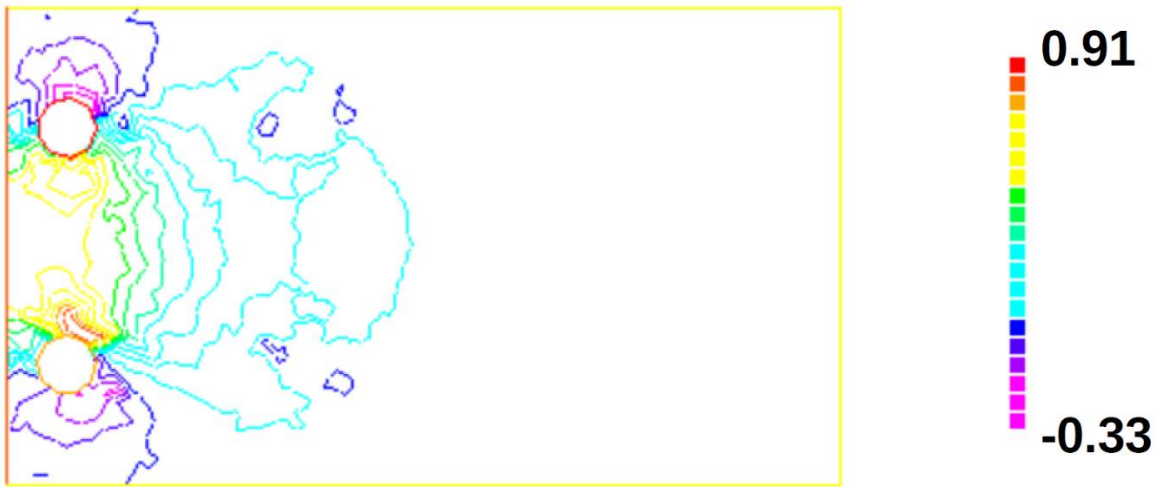

Figure 10: Finite element calculations of the two-point resistance for the chosen contact arrangement. The potential distribution and the distribution of the vertical current density component are shown.

Finite element calculations using the FREEFEM+ software [33] were performed to relate the two-point resistance measured for the chosen contact arrangement to the sheet resistance following the approach of [27] as shown in Figure 10: In the calculations for the chosen sample and contact geometry, potentials of 0 and 1 V were applied to the top and bottom contact, respectively, and the consequent potential distribution was calculated assuming an isotropic sheet resistance of 1 Ω. After that the vertical current density distribution was calculated and integrated along a horizontal line along the sample center to determine the total current between the two contacts. The ratio of the potential difference (of 1 V) and the resulting current corresponds to the two-point resistance. Assuming a negligible contact resistance, the measured two-point resistance $R_{2p}$ corresponds to 0.97 times the sheet resistance $R_S$ ($R_{2p}=0.97*R_S$), which can be expressed by a geometry factor of $W/L=1/0.97=1.03$.

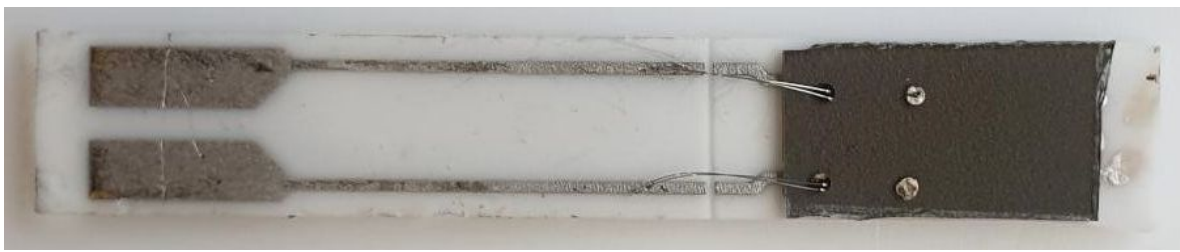

**Figure 11: Photograph of the sample chip with ohmic contacts mounted on a ceramics carrier.**

The chip was subsequently glued with fast-drying silver paint on a ceramics carrier and two of the ohmic contacts were wire-bonded to the contact lines on the carrier to enable in-operando resistance measurements. Figure 11 shows a photograph of the processed sample.

### Operando work function/DC resistance measurements

Two-point DC resistance measurements combined with simultaneously performed work function changes measurements were done at the regular operation temperature of SMOX sensors [12]. Work function changes were measured with the Kelvin Probe technique, which is a non-contact, non-destructive method that uses a vibrating reference electrode and measures the changes of the contact potential difference (CPD) between the sample and the electrode. Variations in the CPD induced by the changes in the gas atmosphere represent relative work function variations of the sample [24]. The sensor, operated at 300°C, was exposed to:

- 5, 10, 20, 50 and 100 ppm of H₂ and CO and 25, 50, 100, 250, 500, 1000 and 2500 ppm of O₂ in a background of N₂;
- 5, 10, 20, 50 and 100 ppm of CO and 2, 5, 10 and 20 ppm of NO₂ in a background of synthetic air.

The total gas flow was 400 sccm. Because the measurements were done in a very dry atmosphere, we assume that the electron affinity remains constant during the whole process [12]. Also, at the working temperature of 300 ºC we assume that gas interactions are limited to the surface. Taking this into account, the changes in CPD upon different gas exposure can be related to work function changes as:

$$\Delta CPD = -\Delta \emptyset = -e\Delta V_s.  \qquad (63)$$

A detailed description of work function and Kelvin Probe working principles can be found in [24].

The dependence of the sample conductance on band bending is shown in Figure 12. In order to obtain a good fit, the experimental points were allowed to move along the direction of the band bending, which means that we took into account the possibility that in N₂ there is a downwards or upwards band bending. The best fit to the model was obtained by considering an upward band bending of 0.22 eV, $G_b = 11$ µS and $L_D/D = 0.115$ as values for the fit parameters. Using these data, we obtained $L_D = 32$ nm for the Debye Length, and $n_b = 3.15 \times 10^{22}$ m⁻³ for the bulk electron concentration – in the calculations we used $\epsilon_r = 12$ as an average value [25].

Consequently, there is an effect, not related to oxygen adsorption, that leads to an upwards band bending in the absence of an oxidizing atmosphere (in pure N₂). A similar situation was found in [26] for undoped polycrystalline SnO₂ where this behavior was attributed to a high oxygen vacancy concentration where some Sn⁺⁴ atoms at the surface change to Sn⁺² and acted as surface electron acceptors. This intrinsic band bending has an influence on the electrical behavior since it affects the concentration of surface species and particularly, see [26], increased the sensitivity in comparison with a sample that showed no intrinsic band bending under the same conditions. Using the formula for bulk conductance $G_b = \frac{e\mu W n_b D}{L}$ and the geometrical factor $W/L=1.03$ of our contact geometry, one obtains an electron mobility of $\mu = 75$ cm²/Vs, which is in good agreement with the (phonon-limited) electron Hall mobility of 60 cm²/Vs measured in single crystalline SnO₂ at 300°C [31].

For the calculated bulk electron concentration, the distance between the bottom of the conduction band and the Fermi level can be estimated from

$$(E_C - E_F) = \frac{k_B T}{e} \ln\left[\frac{N_C}{n_b}\right] \tag{64}$$

Considering that the effective density of states of $N_C = 9.510^{24} \, m^{-3}$ based on a density-of-states effective mass of 0.275 times the free electron mass [30] and a temperature of 573 K, one finds $(E_C - E_F) = 0.285$ eV. All experimental data points are within the validity range of the Boltzmann approximation.

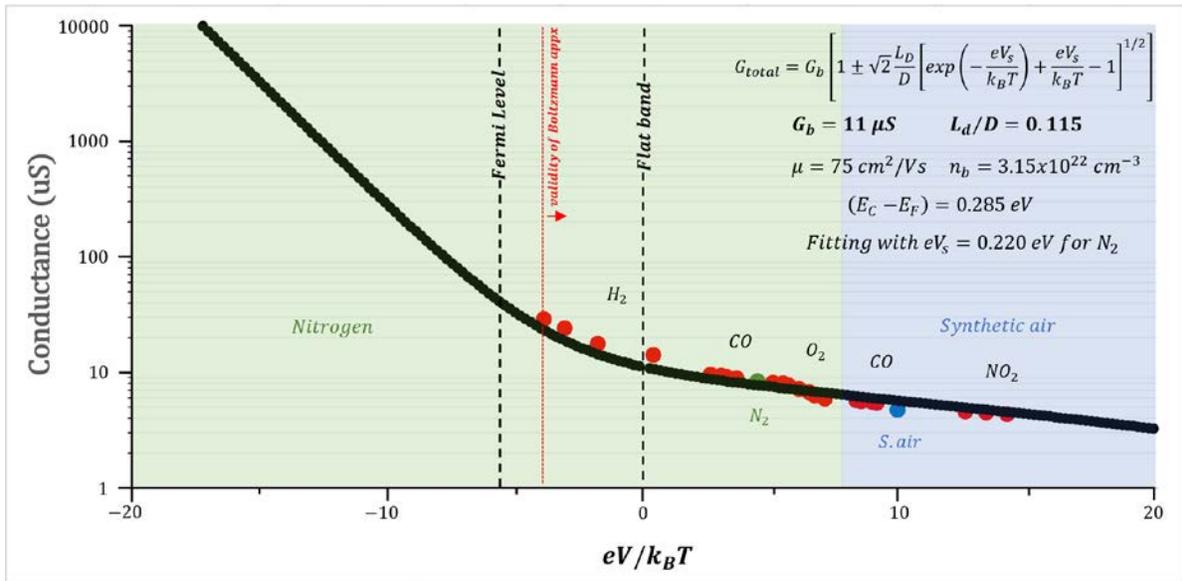

Figure 12: conductance against relative band bending changes. The green and blue area represents the carrier gas present. Allowing experimental points to move along the x direction, a flat band can be found for the first $H_2$ concentration step and, for pure $N_2$, a band bending of 0.22 eV. From the fitting parameters $G_b$ and $L_D/D$, the electronic bulk concentration $n_b = 3.15 \times 10^{22}$ m$^{-3}$ and the mobility $\mu = 75$ cm²/Vs were calculated.

## Conclusion

We developed a theoretical model that correlates the conductance of compact *n* and *p*-type SMOX single crystalline films to the changes in the surface electrostatic potential in the non-degenerate limit. We were also able to apply it to the interpretation of experimental data obtained for an epitaxial $SnO_2(1\,0\,1)$ sensing layer, operated at 300°C and at normal pressure – the investigations were simultaneously performed work function changes and DC conductance measurements. The insights that were gained demonstrate that it is possible to acquire fundamental knowledge that is needed for the basic understanding of gas sensing with SMOX.

**Acknowledgments**

C. E. Simion and Adeline Stanoiu would like to thank the Romanian National Authority for Scientific Research for funding through the Core Program PN19-03 (contract no. 21 N/08.02.2019) and project PN-III-P1-1.1-MC-2017-1917/2017. The work of Federico Schipani was supported by a Georg Forster Fellowship from the Alexander von Humboldt Foundation.